# Unraveling the role of Ta in the phase transition of Pb(Ta$_{1+x}$Se$_2$)$_2$ using low-temperature Raman spectroscopy

Yu Ma[1#], Chi Sin Tang[2#], Xiaohui Yang[3,4#], Yi Wei Ho[5,6#], Jun Zhou[7*], Wenjun Wu[1], Shuo Sun[1], Jin-Ke Bao[8], Dingguan Wang[9], Xiao Lin[10,11], Magdalena Grzeszczyk[5], Shijie Wang[7], Mark B H Breese[2,6], Chuanbing Cai[1], Andrew T. S. Wee[6], Maciej Koperski[5,12*], Zhu-An Xu[4,13,14*], Xinmao Yin[1*]

[1]Shanghai Key Laboratory of High Temperature Superconductors, Institute for Quantum Science and Technology, Department of Physics, Shanghai University, Shanghai 200444, China

[2]Singapore Synchrotron Light Source (SSLS), National University of Singapore, Singapore 117603, Singapore

[3]Department of Physics, China Jiliang University, Hangzhou 310018, Zhejiang, China

[4]School of Physics, Zhejiang University, Hangzhou 310058, China

[5]Institute for Functional Intelligent Materials, National University of Singapore, Singapore 117544, Singapore

[6]Department of Physics, National University of Singapore, Singapore 117551, Singapore

[7]Institute of Materials Research and Engineering (IMRE), Agency for Science, Technology and Research (A*STAR) , 2 Fusionopolis Way, Innovis #08-03, Singapore 138634, Singapore

[8]School of Physics and Hangzhou Key Laboratory of Quantum Matters, Hangzhou Normal University, Hangzhou 311121, China

[9]State Key Laboratory of Radio Frequency Heterogeneous Integration (Shenzhen University), Shenzhen Key Laboratory of Semiconductor Heterogeneous Integration Technology, College of Electronics and Information Engineering, Shenzhen University, Shenzhen 518060, China

[10]Key Laboratory for Quantum Materials of Zhejiang Province, Department of Physics, School of Science and Research Center for Industries of the Future, Westlake University, Hangzhou 310030, China

[11]Institute of Natural Sciences, Westlake Institute for Advanced Study, Hangzhou 310024, China

[12]Department of Materials Science and Engineering, National University of Singapore, Singapore 117575, Singapore

[13]State Key Laboratory of Silicon and Advanced Semiconductor Materials, Zhejiang University, Hangzhou 310027, China

[14]Hefei National Laboratory, Hefei 230088, China

[#]The authors contributed equally to this work.

[*]To whom correspondence should be addressed: zhou_jun@imre.a-star.edu.sg; msemaci@nus.edu.sg; zhuan@zju.edu.cn; yinxinmao@shu.edu.cn

**Abstract:**

Phase engineering strategies in two-dimensional transition metal dichalcogenides (2D-TMDs) have garnered significant attention due to their potential applications in electronics, optoelectronics, and energy storage. Various methods, including direct synthesis, pressure control, and chemical doping, have been employed to manipulate structural transitions in 2D-TMDs. Metal intercalation emerges as an effective technique to modulate phase transition dynamics by inserting external atoms or ions between the layers of 2D-TMDs, altering their electronic structure and physical properties. Here, we investigate the significant structural phase transitions in Pb(Ta$_{1+x}$Se$_2$)$_2$ single crystals induced by Ta intercalation using a combination of Raman spectroscopy and first-principles calculations. The results highlight the pivotal role of Ta atoms in driving these transitions and elucidate the interplay between intercalation, phase transitions, and resulting electronic and vibrational properties in 2D-TMDs. By focusing on Pb(Ta$_{1+x}$Se$_2$)$_2$ as an ideal case study and investigating like metal intercalation, this study advances understanding in the field and paves the way for the development of novel applications for 2D-TMDs, offering insights into the potential of these materials for future technological advancements.

## 1. Introduction:

Phase engineering techniques of two-dimensional transition metal dichalcogenides (2D-TMDs)[1,2] have taken centerstage to better manipulation of their unique properties for potential applications in electronics, optoelectronics, energy storage, and more.[3-5] These structural transitions can be manipulated through diverse methods such as direct synthesis[6-9], pressure control[10-13], and chemical doping[14-18]. For instance, chemical vapor deposition (CVD) enables selective synthesis of single-layer TaSe$_2$ with either trigonal prismatic (1H) or octahedral (1T) crystal structures at varying temperature ranges[6,19]. Multilayer MoS$_2$ exhibits structural distortion at 19 GPa and is followed by a semiconductor-to-metal transition[20]. In addition, doping layered CoSe$_2$ with phosphorus will prompt a structural phase transition of the system from a cubic phase (c-CoSe$_2$) to the orthorhombic (o-CoSe$_2$) phase, thereby creating a catalyst with high hydrogen evolution reaction (HER) activity and stability in alkaline electrolytes[21]. Therefore, the precise engineering of 2D-TMD structural phases is essential to precisely customize their electronic, optical, and mechanical characteristics for respective applications.

The intercalation of metal atoms into 2D-TMDs has been extensively explored as an effective strategy to modulate the system's phase transition dynamics. This process involves the insertion of external atoms or ions between the layers of the 2D-TMDs in consideration, effectively altering the material's electronic structure and physical properties[22-26]. Significant advances have been made in understanding and refining this intercalation technique in 2D-TMDs, widening our knowledge concerning the properties of myriads of 2D-TMD species. For example, V intercalation on quasi-metallic 1T'-phase WSe$_2$ effectively transforms the charge carrier transport characteristics of the material from p-type to n-type[27].

In direct contrast with conventional 2H-phase TaSe$_2$[28,29], its Pb-intercalated counterpart comes in the form of 112-phase PbTaSe$_2$[30]. This manifests itself as a topological superconductor while no longer exhibiting CDW transition.[31,32] The synthesis of the first 124-phase Pb-intercalated Pb(Ta$_{1+x}$Se$_2$)$_2$ is recently reported which exhibits a nontrivial, two-step, reversible, first-order structural phase transition[33]. With these unique transformations of electronic and topological properties induced via metal-intercalation induced phase transition, TaSe$_2$ serves as a model material for elucidating phase transition mechanisms in 2D-TMDs. To investigate how vibrational modes associated with lattice dynamics and electronic excitations would facilitate temperature-dependent structural phase transition processes in single crystal Pb(Ta$_{1+x}$Se$_2$)$_2$, Raman spectroscopy offers unique advantages in characterizing crystalline materials. Nevertheless, it is a regime that remains largely unchartered but plays a pivotal role in validating structural changes and elucidating the influence of Pb atoms and excess Ta atoms[34,35]. By employing Raman spectroscopy, valuable insights could be acquired into the subtle changes taking place in the crystal lattice with temperature, thereby allowing for precise determination of phase transitions and their associated mechanisms[18,36].

Here, we conducted comprehensive experiments integrating Raman spectroscopy measurements with advanced first-principles calculations to investigate significant structural phase transitions in TaSe$_2$ single crystals. Our results demonstrate the substantial influence of Ta atoms in driving these structural transitions, underscoring their pivotal role in the unique behaviors of 2D-TMDs. The electronic and structural properties of Pb(Ta$_{1+x}$Se$_2$)$_2$ is further scrutinized by employing density functional theory (DFT) calculations. Specifically, the energetics of different crystal phases is examined in detail and we validated the structural changes in phases I, II and III induced by Ta intercalation by comparison with temperature-dependent Raman spectra over a wide temperature range down to 4 K [see **Figure 1**(a)]. In Phase I at temperatures above ~250 K, the system is in a structure where the Pb-atom is tetrahedral with a short vertical Pb–Se link alongside three long Pb–Se links. As for structural Phase III at low temperature below ~200 K, the system resembles a linear dumbbell Se–Pb–Se (PbSe$_2$) bonds with short Pb–Se distance of ~3.01 Å. In Phase-II, both tetrahedral and dumbbell-like Pb–Se coordination structure are present in this intermediate phase. This combined experimental-theoretical approach elucidates the mechanisms behind Ta intercalation-induced phase transitions and sheds light on the interplay between intercalation, phase transitions, and the resulting electronic and vibrational properties in 2D-TMDs. By focusing on TaSe$_2$ as a case study and investigating metal intercalation, we highlight the significance of elements such as Pb and Ta in driving structural changes. We aim to advance understanding in the field and facilitate the development of novel applications for 2D TMDs.

## 2. Results:

### 2.1. Sample Information:

Pb(Ta$_{1+x}$Se$_2$)$_2$ single crystals were prepared using the chemical vapor transport (CVT) method with carefully controlled growth temperatures. Pb, Ta, Se powders are mixed according to the ratio 1:2:4 with the addition of PbBr$_2$ as the transport agent. The mixture was then thoroughly mixed and sealed in an evacuated quartz tube with a diameter of 12 mm and length 15 cm. It was then placed in a horizontal two-zone furnace for a week at 900 and 800°C (powders on the warmer side). The resultant single crystals of dimensions ~3 × 3 × 0.2 mm$^3$ were extracted from the middle of the evacuated tube[33].

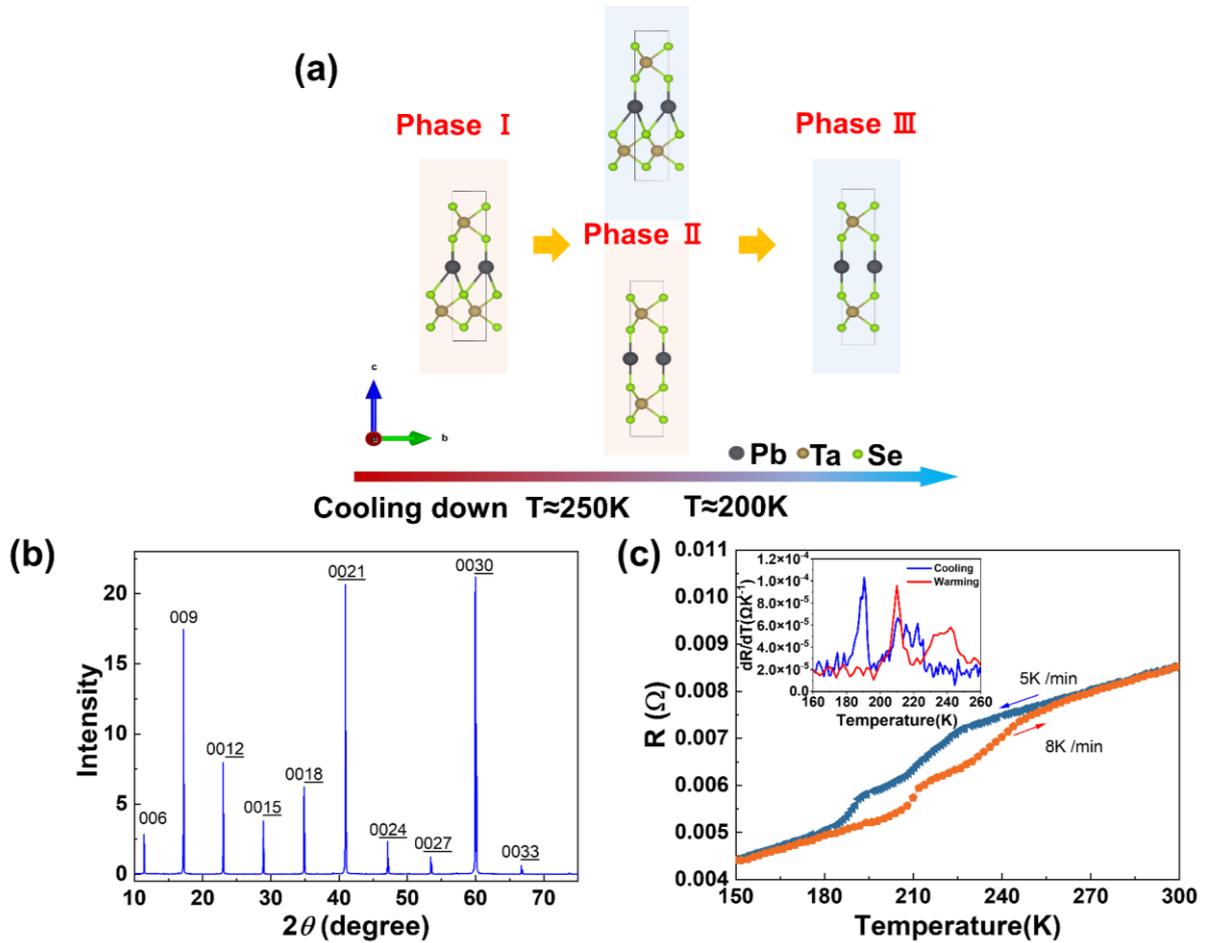

**Figure 1:** Basic properties of Pb(Ta$_{1+x}$Se$_2$)$_2$ single-crystals. (a) Schematic diagram depicting the crystal structure and stacking order with temperature in Phases I, II and III. (b) XRD on the (001) plane of Pb(Ta$_{1+x}$Se$_2$)$_2$ at room temperature,(c) T-dependent electrical resistivity on cooling at a rate 5 K/min$^{-1}$ and warming at a rate 8 K/min$^{-1}$, which shows two step-like anomalies.

The crystalline quality of the Pb(Ta$_{1+x}$Se$_2$)$_2$ single crystal was then assessed via room temperature x-ray diffraction (XRD) ,which was performed on a Bruker D8 Advance x-ray diffractometer with Cu-Kα radiation ,with Figure 1(b) displaying the XRD patterns along the 001-plane of the sample. The XRD spectrum revealed well-defined peaks indexed as (003n) reflections. The result is a clear indication of good crystalline quality, consistent with a previous report[33].

In addition to the sample crystallinity, temperature-dependent resistivity measurements [Figure 1(c)] are conducted to investigate its transport property. The resistance with temperature curve takes place in the range from 150-300 K (Figure 1(c). At ~230 K, the resistivity exhibits two distinct, step-like jumps with noticeable thermal hysteresis. By analyzing the first-order temperature-derivative, [inset of Figure 1(c)], the characteristic temperatures of these consecutive jumps were determined. On warming, the first jump occurs at $T_1 \approx 210$ K, followed by a second jump at $T_2 \approx 242$ K. On cooling, the corresponding temperatures are $T_1 \approx 240$ K and $T_2 \approx 190$ K, respectively. This apparent hysteresis further suggests the first-order nature of the transition[12,18,37,38].

**2.2. Raman spectrum analysis:**

Having verified the quality of the $Pb(Ta_{1+x}Se_2)_2$ sample, temperature-dependent Raman characterization is conducted. The laser wavelength at 532 nm was used to ensure optimal excitation of Raman-active modes present in the intercalated system. This enables a comprehensive analysis of its vibrational spectrum across a wide temperature range. Through a series of temperature-dependent Raman measurements spanning a wide temperature range between 4 and 300 K, changes to the electronic excitations and lattice dynamics can be used to infer the structural phase transition process, from the initial structural changes to the establishment of a thermodynamically stable phase[36,39]. After the acquisition of the Raman spectra, fitting analysis was conducted to extract important quantitative information concerning the lattice parameters, phonon frequencies, and phase fractions, enabling the determination of the phase transition kinetics and thermodynamics. The application of Raman spectroscopy in this study serves as a powerful tool for unraveling the intricate nature of temperature-induced phase transitions in $Pb(Ta_{1+x}Se_2)_2$. Across the temperature range and in different wave number regimes, Raman characteristic peaks were mainly concentrated in the wavenumber ranges of ~ <30 cm$^{-1}$ and ~100 – 150 cm$^{-1}$ [**Figure 2**(a)]. Within wavenumber of Figure 2(a), four distinct Raman features can be identified at ~ 10, ~ 18.6, ~118 cm$^{-1}$ and ~ 136 cm$^{-1}$ where they are labelled *A*, *B*, *C* and *D*, respectively.

Interestingly, as temperature rises, peak *C* remains a single peak (peak *C₁*) at 4 - 200 K (phase

III), but in the temperature range between 200 - 250 K (phase II), a small but clear peak feature (peak $C_2$) near ~120 cm$^{-1}$ appears. As the temperature increases further above 250 K (phase I), peak $C_2$ disappears once again.[Figure (2b)] This phenomenon is consistent with previous studies of the structural phase transition based on temperature-dependent X-ray diffraction of the material[33]. According to the comparison of the optical response of our sample and the substrate, peak *A* and *B* originate from the laser. The detailed information is summarized in Figure S1, Supporting Information. Based on the first-principles calculations, the Raman features in Figure 2(b) can be ascribed to different vibration modes of excess Ta atoms in the material. Details of these vibrational modes and their relationship with the system structural phase transition will be discussed in the next section.

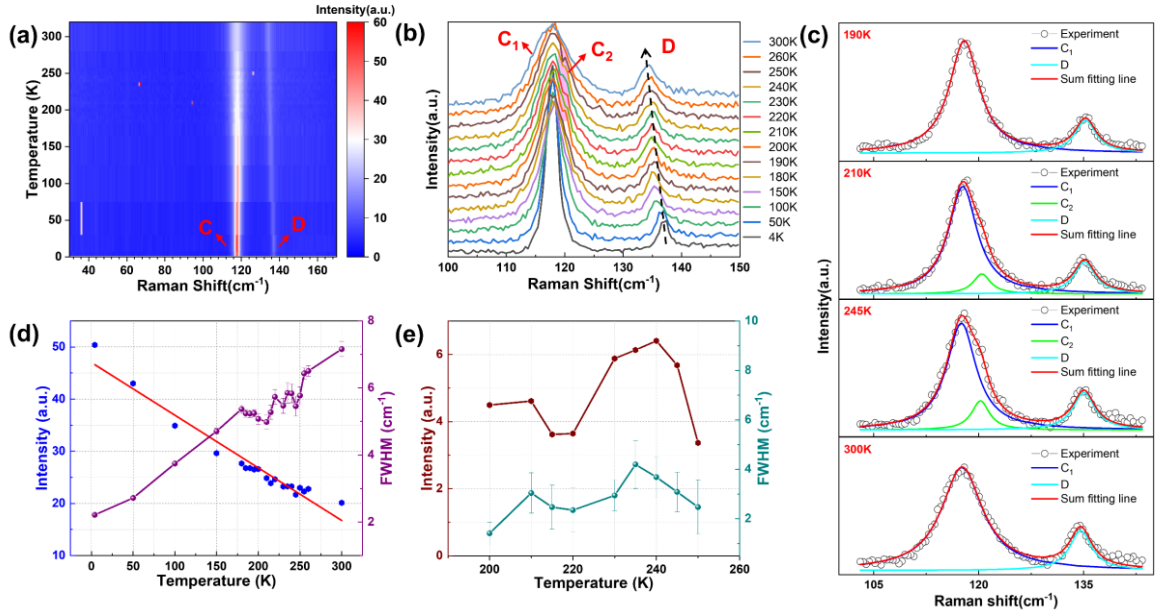

**Figure 2:** Variable-temperature Raman spectrum of Pb(Ta$_{1+x}$Se$_2$)$_2$. (a) Temperature maps of the Raman scattering intensity, which can be seen two Raman peaks (*C* and *D*), (b) Temperature-dependent Raman spectra for peaks *C* and *D*, (c) Curve-fitting of peaks *C* and *D* within 190 K, 210 K, 245 K and 300 K, (d-e) Variation of intensity and FWHM of Raman peak $C_1$ and $C_2$. Other fitted data are shown in the Supporting Information.

Peak fitting of the distinct Raman features is conducted to further analyse and elucidate critical information from the experimental data and to accurately interpret the temperature-dependent Raman spectra due to the nuanced changes observed in their respective intensities and positions. Figure 2(c) shows curve-fitting of peaks *C* and *D* at 190 K, 210 K, 245 K and 300 K, in which the emergence and disappearance of $C_2$ peaks can be seen visually. Detailed analysis reveals significant changes observed for Raman modes *C* and *D* in wavenumber Figure 2(b), where further scrutiny is required. In particular, feature $C_1$ maintains at a consistent position at ~118 cm$^{-1}$ throughout the entire temperature range, thereby indicating a potential stability in the crystal lattice[39]. Despite maintaining a consistent position, the intensity of peak $C_1$ decreases significantly with rising temperature while its Full Width at Half Maximum (FWHM) exhibits a monotonic increase from 2.21 cm$^{-1}$ at 4 K to 7.16 cm$^{-1}$ at 300 K [Figure 2(d)]. In the intermediate temperature range between 200 and 250 K where the structure of Pb(Ta$_{1+x}$Se$_2$)$_2$ is in phase II and where feature $C_2$ emerges, the position of this Raman feature remains largely consistent at ~120 cm$^{-1}$. The intensity and FWHM of this feature do not show any clear temperature dependence with the former varying between 3.36 - 6.4 cm$^{-1}$, while its FWHM between 1.4 - 4.1 cm$^{-1}$.

Raman mode *D* displays a noticeable redshift with rising temperature from ~137 to ~134.6 cm$^{-1}$. This signals an alteration in the lattice vibrational properties. Its intensity is basically stable at about 8 cm$^{-1}$, but there is no obvious linear change for its FWHM, and the overall change range is 2.1 - 3.8 cm$^{-1}$. The details are included in the Supporting Information. Nevertheless, feature *D* has a comparatively lower overall intensity compared to the prominent feature *C* at ~118 cm$^{-1}$. This significant contrast underscores substantial disparities in the information conveyed by features *C* and *D*, respectively, thereby prompting further scrutiny of these features and their relationship, particularly in wavenumber Figure 2(b).

The temperature variations of the Raman features in wavenumber [Figure 2(b)] underscore the dynamic nature of the material phase transition features. This finding is consistent with well-established Raman spectroscopic studies of TMDs. Specifically, the appearance of new Raman features often points to the onset of phase transitions associated with charge density wave (CDW) [36,40,41]. Although previous studies have shown that 3R-TaSe$_2$ has a weak CDW transition, no CDW transition signal has been observed in other previous studies of Pb(Ta$_{1+x}$Se$_2$)$_2$.[33,42] In addition, Raman spectra of the two-phase 2H- and 1T-TaSe$_2$ containing

CDW transitions show that they both have a second-order Raman peak, which is a unique optical feature of CDW materials[36,40]. However, our Raman spectroscopy combined with DFT computational analysis lacks direct evidence of a CDW transition. Instead, it shows a unique structural phase transition due to the change of feature $C_2$. Moreover, the presence of multiple Raman modes observed in Pb(Ta$_{1+x}$Se$_2$)$_2$ and the changes registered with temperature, alongside their temperature-dependent variations, highlight the intricate interactions between different structural phases. Particularly, the appearance of feature $C_2$ exclusively in the intermediate structural phase II challenges initial assumptions regarding a straightforward phase transition mechanism, as clear CDW transition signals are absent. These subtle changes underscore the need for a deeper investigation into the underlying dynamics driving the complex behavior observed in the material, which necessitates further investigation through first-principles calculations for a comprehensive understanding as presented in the next section.

## 2.3. First-principles calculations:

First-principles study is conducted to better understand the phase transition processes of the Pb(Ta$_{1+x}$Se$_2$)$_2$ crystal. By first considering the effects of the interstitial Pb and Ta atoms on the phonon modes, emphasis will be on the wavenumber regions in the vicinity of ~100 cm$^{-1}$ to 150 cm$^{-1}$, respectively based on the above experimental results. As shown in **Figure 3** (a), within these two separate regions, 2H-phase TaSe$_2$ (*P*6$_3$/*mmc*) has Raman active phonon modes of at 13 and 133 cm$^{-1}$, which can be attributed to the $E_{2g}^2$ and $E_{1g}$ modes, respectively. However, with the inclusion of interstitial Pb or Ta atoms to the lattice, the system will be converted into the *P*3*m*1 group space. In which case, it will possess Raman active modes of 6$A_1$+6$E$. In particular, the TaSe$_2$ with Pb (type-1) atoms comprises 3 phonon modes (*E*) with in-plane vibrations at these two separate Raman regions. While the modes arising at 8 cm$^{-1}$ (*E*) and 145 cm$^{-1}$ (*E*) and can correspond to the $E_{2g}^2$-mode at 13 cm$^{-1}$ and the $E_{1g}$-mode at 132 cm$^{-1}$ belonging to 2H-phase TaSe$_2$, an extra phonon mode could be observed arising from the vibration of the included interstitial Pb atoms at around ~ 15 cm$^{-1}$ (*E*). In contrast to type-1 Pb-atoms, type-2 Pb-atoms bond differently to the upper and lower TaSe$_2$ monolayers, and this introduces an asymmetry to the phonon modes. For instance, although the mode at 145 cm$^{-1}$(E) can correspond to the $E_{1g}$-mode at 132 cm$^{-1}$ of 2H-phase TaSe$_2$, the different types of bonding lead to a smaller vibration strength of the Se atoms in the upper TaSe$_2$ layer than that of the lower TaSe$_2$ layer. This results in the emergence of the new vibration patterns between Pb and TaSe$_2$ emerge at both 10 (E) and 18 cm$^{-1}$(E) [refer to Fig. 3(a)].

The inclusion of Ta atoms also breaks the symmetry of the two TaSe$_2$ monolayers. This is because the interstitial Ta atoms form stronger bonds with the lower Se atoms (bond length: 2.61 Å) than that of the upper Se atoms (2.75 Å). This leads to a significantly stronger vibration of the Se atoms in the upper section of the TaSe$_2$ monolayer which manifests itself at 113 cm$^{-1}$ (E). The introduction of the interstitial Ta atoms also induces a new vibration mode at ~124 cm$^{-1}$ ($A_1$) that is primarily attributed to the out-of-plane vibration of the interstitial Ta atoms. However, there is no phonon frequency at the wavenumber region around 10 cm$^{-1}$ induced by interstitial Ta doping [Fig 3(a) right schematic].

These calculated results suggest the necessary inclusion of the synergetic effects of interstitial Ta atoms to explain the experimentally observed Raman spectra. As we compare the phonon modes of 2H- phase TaSe$_2$ and the three different structural phases with Pb doping as well as the co-doping of both Pb and Ta atoms, we find that although there are phonon modes in wavenumber region around 10 cm$^{-1}$, there are also no phonon modes within the region in the range of 118 and 136 cm$^{-1}$ [Figure 3 (b)]. It is only by including the interstitial Ta atoms that can lead to the emergence of phonon modes at both of these wavenumber regions. These results suggest the existence of interstitial Ta atoms in the Pb(Ta$_{1+x}$Se$_2$)$_2$ samples.

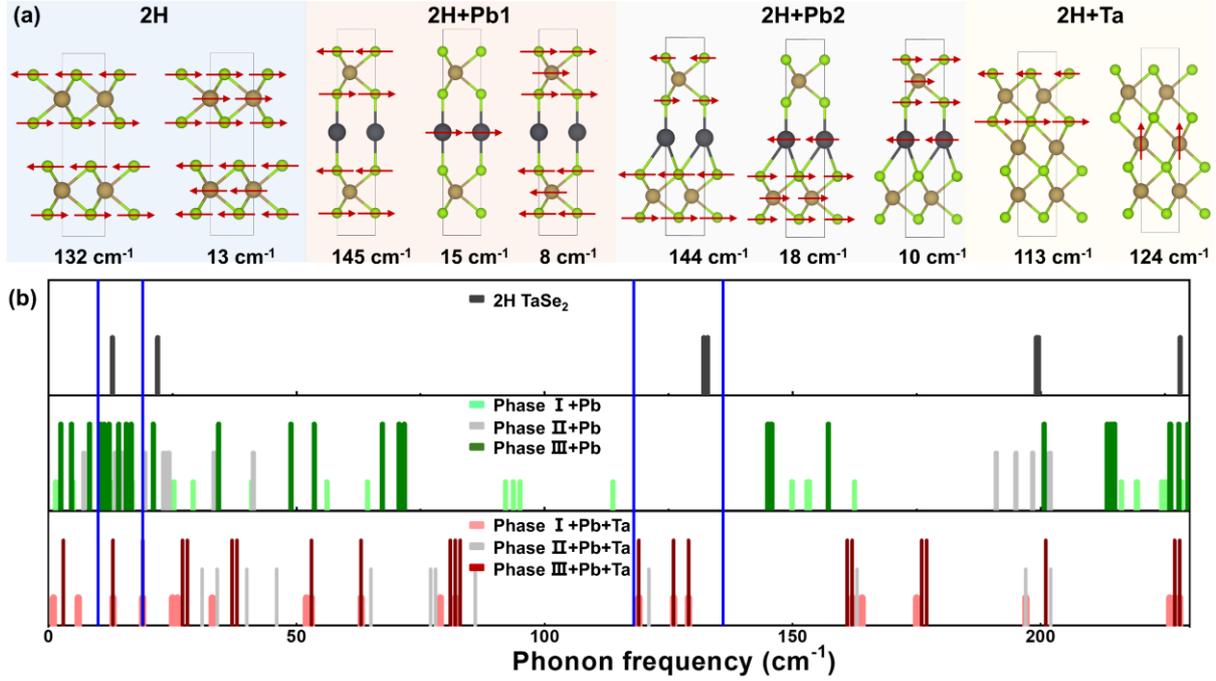

**Figure 3:** DFT calculation of 2H-TaSe$_2$ Raman spectra. (a) The effects of the interstitial Ta and Pb atoms on the phonon frequencies of 2H TaSe$_2$. (b) The phonon frequencies of 2H TaSe$_2$, phase I, II and III with interstitial Pb atoms, as well as phase I, II and III with both interstitial Pb and Ta atoms.

To further examine the structural phase transition properties of the Pb(Ta$_{1+x}$Se$_2$)$_2$ sample, attention will be on the interstitial Ta and Pb co-doped phases. It is noted that the structures used in the DFT calculations assume a perfectly ordered bulk crystal at 0 K, and therefore extrinsic factors associated with temperature, pressure, defects, and phonon anharmonicity are not considered. According to high-throughput DFT simulations of Raman spectra and statistical analysis with experimental data [43,44], the average wave number deviation and standard deviation is −9.66 cm$^{-1}$ and 18.58 cm$^{-1}$, respectively. The key structural parameters are listed in the Supporting Information of Table S1. Besides, DFT tend to exhibit larger errors at lower wavenumbers. i.e., these modes correspond to long-range oscillations that are difficult to compute within the limited size of periodic unit cells. Thus, quantitative comparison between experimentally derived Raman spectra and theoretical results are difficult. DFT calculations yielded Raman active modes in the regions ~119 - 120 cm$^{-1}$ and 128 - 129 cm$^{-1}$. These fall within the average deviation for experimentally observed Raman modes $C_1$, $C_2$, and $D$ at ~118 and ~136 cm$^{-1}$, respectively. In contrast to the calculations, no modes around 126 cm$^{-1}$ were seen in the Raman spectra. This discrepancy may be due to the low intensity of this vibrational modes.[45] To compensate for the intrinsic inaccuracy of DFT simulations, the Raman modes have been shifted with same vibration patterns for the three phases to a single wavenumber.

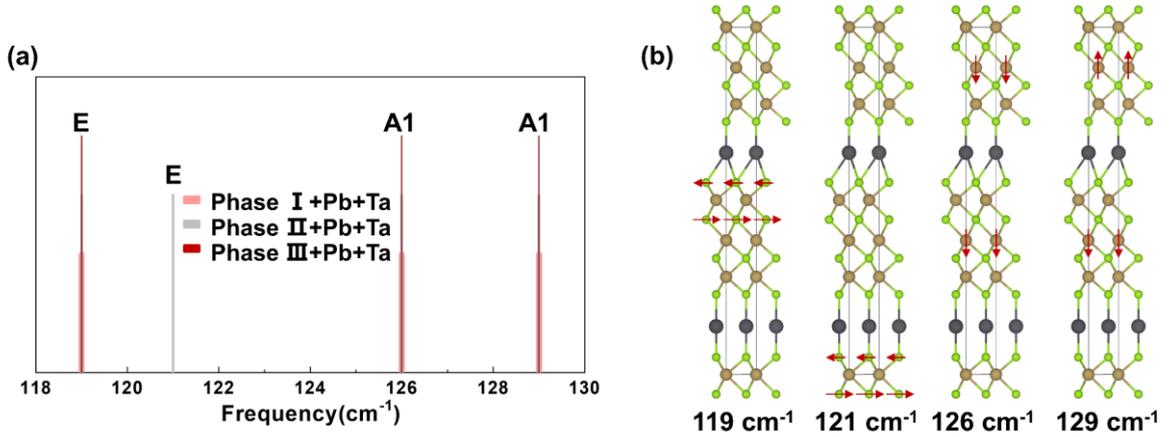

**Figure 4**: DFT calculation of Pb(Ta$_{1+x}$Se$_2$)$_2$ Raman spectra. (a) Different phonon modes within the 118-130 cm$^{-1}$ region. (b) The effects of the interstitial Ta atoms on the phonon frequencies of phase II in (a).

As a result, four different phonon modes can be observed in the wavenumber region at ~118-136 cm$^{-1}$ [**Figure 4** (a)]. Their corresponding vibration patterns are displayed in Figure 4(b), where the phonons of phase II are displayed for simplicity (See supporting information for the vibration patterns for phases I and III). At ~119 cm$^{-1}$, the phonon mode is corresponding to an in-plane vibration of the two Se atom layers with opposite direction bonded to type-2 Pb atoms. As for the Raman mode at ~121 cm$^{-1}$, similar Se vibrations take place but these Se atoms are bonded with type-1 Pb atoms. In the case of the Raman mode at ~126 cm$^{-1}$, it corresponds to the two interstitial layers of Ta atoms that vibrate vertically in the same directions. Meanwhile, the mode at ~129 cm$^{-1}$ can be attributed to similar interstitial Ta atoms contributed vibrations but with the two interstitial Ta atoms vibrating along opposite directions. At the low wavenumber region at ~10 - 19 cm$^{-1}$, we also got the modes at ~13 and ~19 cm$^{-1}$ by the DFT calculations, which are mainly the results of in-plane vibrations that arise due to type-1 and type-2 Pb atoms, respectively (see supporting information, Figure S4). It may be due to the influence of the instrument laser, so we cannot accurately extract this part of the peak in the low frequency region.

With the origins of the respective Raman modes established, we propose a model to explain our temperature-dependent experimental results. In the low-temperature regime (lower than ~200 K), phase III Pb(Ta$_{1+x}$Se$_2$)$_2$ structure consists of type-1 Pb atoms, which manifests itself with the presence of phonons modes at ~119 and ~129 cm$^{-1}$. There is a structural transition from phase III to phase II with rising temperature due to the two types of interstitial Pb atoms. This results in two classes of phonon modes at 119 and 121 cm$^{-1}$. At even higher temperature (higher than ~250 K), structural transition will take place from phase II to phase I. Similar to the low-temperature phase III, phase I only comprises a type of Pb atoms. Therefore, there is only a single class of phonons at ~119 cm$^{-1}$.

As the temperature increases, the Raman spectra exhibit a distinct one-two-one peak transition at approximately 119 cm$^{-1}$ (peak $C_1$ and $C_2$), indicative of the two-fold structural transition process. Notably, phonon modes around 119 cm$^{-1}$ primarily involve in-plane vibrations, rendering them less sensitive to temperature-induced lattice parameter changes and resulting in negligible peak shifts in this region. In contrast, vibrations at 129 cm$^{-1}$ primarily involve out-of-plane motions, rendering them more susceptible to temperature variations. Consequently, the observed red shift in the Raman spectra for peak $D$ can be attributed to the heightened sensitivity of these vibrations to temperature increases.

## 3. Conclusion:

In this study, we performed first-principles calculations to gain further insights into the Raman spectroscopic features and phase transition behavior of Pb(Ta$_{1+x}$Se$_2$)$_2$. The computed Raman peak positions for the three different phases (phase I, phase II, and phase III) slightly deviated from our experimental measurements, which can be attributed to the inherent inaccuracies in the calculations.

The Raman spectroscopic analysis confirmed the presence of a structural phase transition in the material, which varies with temperature. The emergence of new Raman peaks in the 200-250K range further supported this conclusion. The calculations were consistent with our experimental observations, validating the accuracy of our Raman fitting results.

Further investigation into the vibrations associated with the *C* and *D* peaks provided valuable insights. It was found that the *C* peak is attributed to in-plane transverse vibrations of the pristine TaSe$_2$ plane, which explains why they do not show a significant shift with temperature variations. On the other hand, the intensity variation and pronounced redshift of the *D* peak with increasing temperature were attributed to vibrations of excess tantalum atoms in the lattice, which also accounts for its relatively lower intensity compared to the *C* peak.

Additionally, our calculations observed lower-frequency and lower-intensity Raman active peaks around 10 cm$^{-1}$ and 20 cm$^{-1}$, which were attributed to interlayer vibrations of the lead atoms. Unfortunately, due to the limitation of instruments, we have not obtained accurate experimental verification, and we need to further explore the Raman spectrum in the low frequency region.

Overall, our first-principles calculations and vibrational analysis provided a comprehensive understanding of the Raman spectroscopic features and phase transition in Pb(Ta$_{1+x}$Se$_2$)$_2$. The consistency between the computational results and experimental observations validates the reliability of our approach. These findings contribute to a broader understanding of the complex phase transition mechanisms in this material. Future research can build upon these results to explore the effects of different parameters such as strain or doping on the Raman spectroscopy and phase transitions of Pb(Ta$_{1+x}$Se$_2$)$_2$.

**Acknowledgements**
This work was supported by National Natural Science Foundation of China (Grant No.52172271, 12374378, 52307026), the National Key R&D Program of China (Grant No. 2022YFE03150200), Shanghai Science and Technology Innovation Program (Grant No. 22511100200, 23511101600).
C.S.T acknowledges the support from the NUS Emerging Scientist Fellowship. Z. X.

acknowledges the support from the National Science Foundation of China (Grant No. 12174334) and the National Key Projects for Research & Development of China (Grant No. 2019YFA0308602).

Xiaohui Yang acknowledges the support from Zhejiang Provincial Natural Science Foundation of China (Grant No. LQ23A040009) and National Natural Science Foundation of China (Grants No. 12304168).

Yi Wei Ho acknowledges the support from the Ministry of Education (MOE), Singapore, under Academic Research Fund (AcRF) Tier 3 (Grant No. MOE2018-T3-1-005).

Xiao Lin acknowledges the support from Zhejiang Provincial Natural Science Foundation of China for Distinguished Young Scholars (Grant No. LR23A040001).

Jun Zhou acknowledges the computational resources supported by the National Supercomputing Centre (NSCC) Singapore.

**Competing interests**

The authors declare no competing interests.

**References**

[1]   H. H. Huang, X. Fan, D. J. Singh, W. T. Zheng. Nanoscale. 2020, 12, 1247-1268.
[2]   R. Wang, Y. Yu, S. Zhou, H. Li, H. Wong, Z. Luo, L. Gan, T. Zhai. Advanced Functional Materials. 2018, 28, 1802473.
[3]   C. Gong, Y. Zhang, W. Chen, J. Chu, T. Lei, J. Pu, L. Dai, C. Wu, Y. Cheng, T. Zhai, L. Li, J. Xiong. Advanced Science. 2017, 4, 1700231.
[4]   J. Ji, J. H. Choi. Nanoscale. 2022, 14, 10648-10689.
[5]   T. Mueller, E. Malic. npj 2D Materials and Applications. 2018, 2, 29.
[6]   K.-A. N. Duerloo, Y. Li, E. J. Reed. Nature Communications. 2014, 5, 4214.
[7]   Y.-C. Lin, D. O. Dumcenco, Y.-S. Huang, K. Suenaga. Nature Nanotechnology. 2014, 9, 391-396.
[8]   M. Liu, J. Gou, Z. Liu, Z. Chen, Y. Ye, J. Xu, X. Xu, D. Zhong, G. Eda, A. T. S. Wee. Nature Communications. 2024, 15, 1765.
[9]   A. Ohtake, X. Yang, J. Nara. npj 2D Materials and Applications. 2022, 6, 35.
[10]  P. Johari, V. B. Shenoy. ACS Nano. 2012, 6, 5449-5456.
[11]  T. Lin, X. Wang, X. Chen, X. Liu, X. Luo, X. Li, X. Jing, Q. Dong, B. Liu, H. Liu, Q. Li, X. Zhu, B. Liu. Inorganic Chemistry. 2021, 60, 11385-11393.
[12]  Y. Qi, P. G. Naumov, M. N. Ali, C. R. Rajamathi, W. Schnelle, O. Barkalov, M. Hanfland, S.-C. Wu, C. Shekhar, Y. Sun, V. Süß, M. Schmidt, U. Schwarz, E. Pippel, P. Werner, R. Hillebrand, T. Förster, E. Kampert, S. Parkin, R. J. Cava, C. Felser, B.

Yan,S. A. Medvedev. Nature Communications. 2016, 7, 11038.

[13] Y. Xiao, S. He, M. Li, W. Sun, Z. Wu, W. Dai,C. Lu. Scientific Reports. 2021, 11, 22090.

[14] D. Han, W. Ming, H. Xu, S. Chen, D. Sun,M.-H. Du. Physical Review Applied. 2019, 12, 034038.

[15] W. Li, J. Huang, B. Han, C. Xie, X. Huang, K. Tian, Y. Zeng, Z. Zhao, P. Gao, Y. Zhang, T. Yang, Z. Zhang, S. Sun,Y. Hou. Advanced Science. 2020, 7, 2001080.

[16] Y.-C. Lin, R. Torsi, D. B. Geohegan, J. A. Robinson,K. Xiao. Advanced Science. 2021, 8, 2004249.

[17] T. Zhang, K. Fujisawa, F. Zhang, M. Liu, M. C. Lucking, R. N. Gontijo, Y. Lei, H. Liu, K. Crust, T. Granzier-Nakajima, H. Terrones, A. L. Elías,M. Terrones. ACS Nano. 2020, 14, 4326-4335.

[18] Y. Zhang, F. Fei, R. Liu, T. Zhu, B. Chen, T. Qiu, Z. Zuo, J. Guo, W. Tang, L. Zhou, X. Xi, X. Wu, D. Wu, Z. Zhong, F. Song, R. Zhang,X. Wang. Advanced Materials. 2023, 35, 2207841.

[19] Y. Chen, W. Ruan, M. Wu, S. Tang, H. Ryu, H.-Z. Tsai, R. L. Lee, S. Kahn, F. Liou, C. Jia, O. R. Albertini, H. Xiong, T. Jia, Z. Liu, J. A. Sobota, A. Y. Liu, J. E. Moore, Z.-X. Shen, S. G. Louie, S.-K. Mo,M. F. Crommie. Nature Physics. 2020, 16, 218-224.

[20] A. P. Nayak, S. Bhattacharyya, J. Zhu, J. Liu, X. Wu, T. Pandey, C. Jin, A. K. Singh, D. Akinwande,J.-F. Lin. Nature Communications. 2014, 5, 3731.

[21] Y.-R. Zheng, P. Wu, M.-R. Gao, X.-L. Zhang, F.-Y. Gao, H.-X. Ju, R. Wu, Q. Gao, R. You, W.-X. Huang, S.-J. Liu, S.-W. Hu, J. Zhu, Z. Li,S.-H. Yu. Nature Communications. 2018, 9, 2533.

[22] R. Haverkamp, N. L. A. N. Sorgenfrei, E. Giangrisostomi, S. Neppl, D. Kühn,A. Föhlisch. Scientific Reports. 2021, 11, 6893.

[23] P. Kumar, R. Skomski,R. Pushpa. ACS Omega. 2017, 2, 7985-7990.

[24] F. Wang, Y. Zhang, Z. Wang, H. Zhang, X. Wu, C. Bao, J. Li, P. Yu,S. Zhou. Nature Communications. 2023, 14, 4945.

[25] X. Zhao, P. Song, C. Wang, A. C. Riis-Jensen, W. Fu, Y. Deng, D. Wan, L. Kang, S. Ning, J. Dan, T. Venkatesan, Z. Liu, W. Zhou, K. S. Thygesen, X. Luo, S. J. Pennycook,K. P. Loh. Nature. 2020, 581, 171-177.

[26] X.-C. Liu, S. Zhao, X. Sun, L. Deng, X. Zou, Y. Hu, Y.-X. Wang, C.-W. Chu, J. Li, J. Wu, F.-S. Ke,P. M. Ajayan. Science Advances. 2020, 6, eaay4092.

[27] K. Zhu, Y. Tao, D. E. Clark, W. Hong,C. W. Li. Nano Letters. 2023, 23, 4471-4478.

[28] J. A. Wilson, F. J. Di Salvo,S. Mahajan. Physical Review Letters. 1974, 32, 882-885.

[29] K.-i. Yokota, G. Kurata, T. Matsui,H. Fukuyama. Physica B: Condensed Matter. 2000, 284-288, 551-552.

[30] R. Eppinga,G. A. Wiegers. Physica B+C. 1980, 99, 121-127.

[31] G. Bian, T.-R. Chang, R. Sankar, S.-Y. Xu, H. Zheng, T. Neupert, C.-K. Chiu, S.-M. Huang, G. Chang, I. Belopolski, D. S. Sanchez, M. Neupane, N. Alidoust, C. Liu, B. Wang, C.-C. Lee, H.-T. Jeng, C. Zhang, Z. Yuan, S. Jia, A. Bansil, F. Chou, H. Lin,M. Z. Hasan. Nature Communications. 2016, 7, 10556.

[32] S.-Y. Guan, P.-J. Chen, M.-W. Chu, R. Sankar, F. Chou, H.-T. Jeng, C.-S. Chang,T.-M. Chuang. Science Advances. 2016, 2, e1600894.

[33] X. Yang, J.-K. Bao, Z. Lou, P. Li, C. Jiang, J. Wang, T. Sun, Y. Liu, W. Guo, S.

Ramakrishnan, S. R. Kotla, M. Tolkiehn, C. Paulmann, G.-H. Cao, Y. Nie, W. Li, Y. Liu, S. van Smaalen, X. Lin,Z.-A. Xu. Advanced Materials. 2022, 34, 2108550.

[34] O. R. Albertini, R. Zhao, R. L. McCann, S. Feng, M. Terrones, J. K. Freericks, J. A. Robinson,A. Y. Liu. Physical Review B. 2016, 93, 214109.

[35] A. A. Puretzky, L. Liang, X. Li, K. Xiao, K. Wang, M. Mahjouri-Samani, L. Basile, J. C. Idrobo, B. G. Sumpter, V. Meunier,D. B. Geohegan. ACS Nano. 2015, 9, 6333-6342.

[36] H. Wang, Y. Chen, C. Zhu, X. Wang, H. Zhang, S. H. Tsang, H. Li, J. Lin, T. Yu, Z. Liu,E. H. T. Teo. Advanced Functional Materials. 2020, 30, 2001903.

[37] G. Jarc, S. Y. Mathengattil, A. Montanaro, F. Giusti, E. M. Rigoni, R. Sergo, F. Fassioli, S. Winnerl, S. Dal Zilio, D. Mihailovic, P. Prelovšek, M. Eckstein,D. Fausti. Nature. 2023, 622, 487-492.

[38] Y. Yu, F. Yang, X. F. Lu, Y. J. Yan, Y.-H. Cho, L. Ma, X. Niu, S. Kim, Y.-W. Son, D. Feng, S. Li, S.-W. Cheong, X. H. Chen,Y. Zhang. Nature Nanotechnology. 2015, 10, 270-276.

[39] A. S. Sarkar,S. K. Pal. ACS Omega. 2017, 2, 4333-4340.

[40] P. Hajiyev, C. Cong, C. Qiu,T. Yu. Scientific Reports. 2013, 3, 2593.

[41] D. Lin, S. Li, J. Wen, H. Berger, L. Forró, H. Zhou, S. Jia, T. Taniguchi, K. Watanabe, X. Xi,M. S. Bahramy. Nature Communications. 2020, 11, 2406.

[42] Y. Deng, Y. Lai, X. Zhao, X. Wang, C. Zhu, K. Huang, C. Zhu, J. Zhou, Q. Zeng, R. Duan, Q. Fu, L. Kang, Y. Liu, S. J. Pennycook, X. R. Wang,Z. Liu. Journal of the American Chemical Society. 2020, 142, 2948-2955.

[43] M. Bagheri,H. P. Komsa. Sci Data. 2023, 10, 80.

[44] Q. Liang, S. Dwaraknath,K. A. Persson. Sci Data. 2019, 6, 135.

[45] N. Fleck, T. D. C. Hobson, C. N. Savory, J. Buckeridge, T. D. Veal, M. R. Correia, D. O. Scanlon, K. Durose,F. Jäckel. Journal of Materials Chemistry A. 2020, 8, 8337-8344.

# Supporting Information

**Unraveling the role of Ta in the phase transition of Pb(Ta$_{1+x}$Se$_2$)$_2$ using Low-Temperature Raman spectroscopy**

Yu Ma[#], Chi Sin Tang[#], Xiaohui Yang[#], Yi Wei Ho[#], Jun Zhou[*], Wenjun Wu, Shuo Sun, Jin-Ke Bao, Dingguan Wang, Xiao Lin, Magdalena Grzeszczyk, Shijie Wang, Mark B H Breese, Chuanbing Cai, Andrew T. S. Wee, Maciej Koperski[*], Zhu-An Xu[*], Xinmao Yin[*]

[#] These authors contributed equally to this work.
[*] Corresponding author. E-mail: zhou_jun@imre.a-star.edu.sg; msemaci@nus.edu.sg; zhuan@zju.edu.cn; yinxinmao@shu.edu.cn

The PDF file includes:

Section S1. Detailed information of growth conditions

Section S2. Data processing in Raman Spectroscopy

Section S3. DFT Calculations

Figure S1. Variable-temperature Raman spectrum off the sample

Figure S2. Curve-fitting of peak D between 4K to 300K

Figure S3. The vibration patterns of Pb(Ta$_{1+x}$Se$_2$)$_2$ for phases I and III

Figure S4. DFT calculation of Pb(Ta$_{1+x}$Se$_2$)$_2$ Raman spectra for phase II

Table S1. The relaxed structures of Pb-intercalated TaSe$_2$.

**Section S1. Detailed information of growth conditions**

Pb(Ta$_{1+x}$Se$_2$)$_2$ single crystal was prepared by chemical vapor transport (CVT) method. Pb, Ta and Se powders were mixed at a ratio of 1:2:4, and PbBr$_2$ was added as a transport agent with a concentration about 10 mg/cm$^3$. This was followed by a vacuum seal in a quartz tube with a diameter of 12 mm and a length of 15 cm, in a horizontal double-temperature zone tube furnace with a set temperature of 900°C and 800°C (powder on the hotter side). The temperature gradient is 9 °C/cm, and the single crystal is finally crystallized at a distance of 8 cm from the raw material, and the extraction size is ~3×3× 0.2mm$^3$.[1]

**Section S2. Data processing in Raman Spectroscopy**

A 532 nm CW laser (Cobolt 08-01) filtered with an interference filter (Semrock LL-01-532) and a reflective Bragg grating bandpass filter (OptiGrate BP-532) before illuminate on the sample mounted inside a closed-cycle cryostat (attocube attoDRY800), focused through a cold objective (attocube LT-APO/VIS/0.82). The reflected signal is collected with the same lens and split by a non-polarizing beam splitter (Thorlabs BS013), filtered with 3 Gragg grating notch filters (OptiGrate BNF-532) before measured with a spectrometer (Princeton Instruments, 1800 g/mm grating and PyLoN 100BR eXcelon CCD). Raman spectra were acquired with sample temperature ranging from 4 to 300 K with various temperature interval, at each temperature the sample was adjusted to ensure the laser was focused and signal was collected on the same spot on the flake.

In order to eliminate the influence of some interfering components, such as substrate or laser, we removed the sample and performed Raman spectroscopy with temperature change. It turns out that peaks *A* and *B* are not from the sample itself.

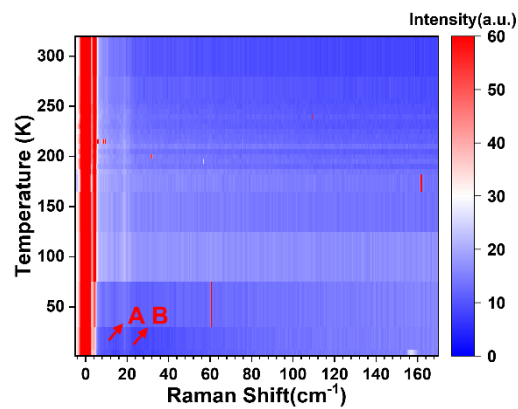

**Figure S1. Variable-temperature Raman spectrum off the sample.** Temperature maps of the Raman scattering intensity, which can be seen two Raman peaks (peak *A* and *B*).

Figure S2 is a supplement to our fitting results for peak *D*. It can be clearly seen that peak *D* shows an obvious redshift with the increase of temperature, while its intensity is stable at about 8cm$^{-1}$.

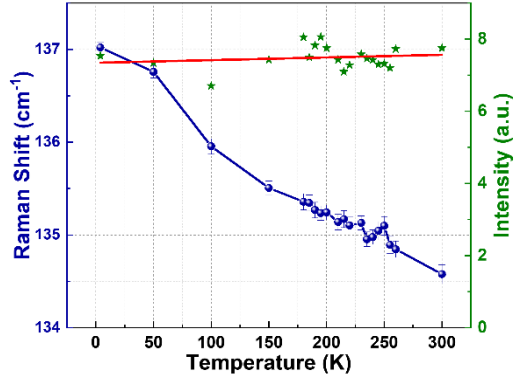

**Figure S2. Curve-fitting of peak *D* between 4K to 300K.** Variation of Raman Shift and intensity of Raman peak *D*.

**Section S3. DFT Calculations**

The first-principles calculations were conducted employing Vienna *ab initio* Simulation Package (VASP) based on density functional theory [2,3]. The Perdew-Burke-Ernzerhof (PBE) approximation was adopted for the exchange-correlation functional, while the electron-ion interaction was treated using the frozen-core all-electron projector augmented wave (PAW) method [4]. DFT-D3 method with Becke-Johnson damping function was applied to estimate the van der Waals (vdW) dispersion interactions for layered structures [5]. The cutoff energy for the plane wave expansion was set to 600 eV. Γ-centered 16×16×1 *k*-meshes are used. The structures were fully relaxed until the energy and force were converged to 10$^{-5}$ eV and 0.01 eV/Å, respectively. The phonon modes are calculated density-functional perturbation theory [6]. The point group and the Raman-active irreducible representation of the modes are extracted from the Bilbao Crystallographic Server [7].

The structural parameters of the three phases are shown in Table S1.

**Table S1.** The relaxed structures of Pb-intercalated TaSe$_2$.

| Phase I | | | | Phase II | | | | Phase III | | | |
|---|---|---|---|---|---|---|---|---|---|---|---|
| a = b = 3.345 Å; c = 50.154 Å | | | | a = b = 3.343 Å; c = 33.633 Å | | | | a = b = 3.351 Å; c = 50.765 Å | | | |
| α = β = 90°, γ = 120° | | | | α = β = 90°, γ = 120° | | | | α = β = 90°, γ = 120° | | | |
| Atom | x | y | z | Atom | x | y | z | Atom | x | y | z |
| Ta1 | 0.000 | 0.000 | 0.000 | Ta1 | 0.000 | 0.000 | 0.000 | Ta1 | 0.000 | 0.000 | 0.000 |
| Ta2 | 0.333 | 0.667 | 0.185 | Ta2 | 0.667 | 0.333 | 0.779 | Ta2 | 0.333 | 0.667 | 0.520 |
| Ta3 | 0.000 | 0.000 | 0.519 | Ta3 | 0.667 | 0.333 | 0.884 | Ta3 | 0.000 | 0.000 | 0.187 |
| Ta4 | 0.667 | 0.333 | 0.334 | Ta4 | 0.000 | 0.000 | 0.281 | Ta4 | 0.667 | 0.333 | 0.667 |
| Ta5 | 0.667 | 0.333 | 0.852 | Ta5 | 0.333 | 0.667 | 0.502 | Ta5 | 0.667 | 0.333 | 0.853 |
| Ta6 | 0.333 | 0.667 | 0.667 | Ta6 | 0.000 | 0.000 | 0.386 | Ta6 | 0.333 | 0.667 | 0.333 |
| Ta7 | 0.667 | 0.333 | 0.922 | Se1 | 0.000 | 0.000 | 0.832 | Ta7 | 0.667 | 0.333 | 0.923 |
| Ta8 | 0.000 | 0.000 | 0.589 | Se2 | 0.000 | 0.000 | 0.728 | Ta8 | 0.000 | 0.000 | 0.256 |
| Ta9 | 0.333 | 0.667 | 0.256 | Se3 | 0.333 | 0.667 | 0.050 | Ta9 | 0.333 | 0.667 | 0.590 |
| Se1 | 0.667 | 0.333 | 0.221 | Se4 | 0.333 | 0.667 | 0.943 | Se1 | 0.667 | 0.333 | 0.555 |
| Se2 | 0.333 | 0.667 | 0.962 | Se5 | 0.333 | 0.667 | 0.335 | Se2 | 0.333 | 0.667 | 0.963 |
| Se3 | 0.333 | 0.667 | 0.034 | Se6 | 0.333 | 0.667 | 0.231 | Se3 | 0.333 | 0.667 | 0.033 |
| Se4 | 0.667 | 0.333 | 0.151 | Se7 | 0.667 | 0.333 | 0.553 | Se4 | 0.667 | 0.333 | 0.487 |
| Se5 | 0.333 | 0.667 | 0.485 | Se8 | 0.667 | 0.333 | 0.446 | Se5 | 0.333 | 0.667 | 0.153 |
| Se6 | 0.333 | 0.667 | 0.554 | Pb1 | 0.000 | 0.000 | 0.639 | Se6 | 0.333 | 0.667 | 0.222 |
| Se7 | 0.000 | 0.000 | 0.888 | Pb2 | 0.333 | 0.667 | 0.139 | Se7 | 0.000 | 0.000 | 0.889 |
| Se8 | 0.000 | 0.000 | 0.296 | | | | | Se8 | 0.000 | 0.000 | 0.629 |
| Se9 | 0.667 | 0.333 | 0.701 | | | | | Se9 | 0.667 | 0.333 | 0.366 |
| Se10 | 0.000 | 0.000 | 0.818 | | | | | Se10 | 0.000 | 0.000 | 0.820 |
| Se11 | 0.000 | 0.000 | 0.367 | | | | | Se11 | 0.000 | 0.000 | 0.700 |
| Se12 | 0.667 | 0.333 | 0.629 | | | | | Se12 | 0.667 | 0.333 | 0.296 |
| Pb1 | 0.667 | 0.333 | 0.091 | | | | | Pb1 | 0.667 | 0.333 | 0.425 |
| Pb2 | 0.000 | 0.000 | 0.758 | | | | | Pb2 | 0.000 | 0.000 | 0.759 |
| Pb3 | 0.333 | 0.667 | 0.425 | | | | | Pb3 | 0.333 | 0.667 | 0.092 |

The vibration patterns of Pb(Ta$_{1+x}$Se$_2$)$_2$ for phases I and III are shown in figure S3. They correspond to the calculation results of the peaks of Raman spectrum. Similarly, the vibration of Pb atom mainly affects the Raman peak at 0-30cm$^{-1}$, while the vibration of Ta atom mainly affects the Raman peak at higher frequencies. The only difference with the phase II is the disappearance of $C_2$.

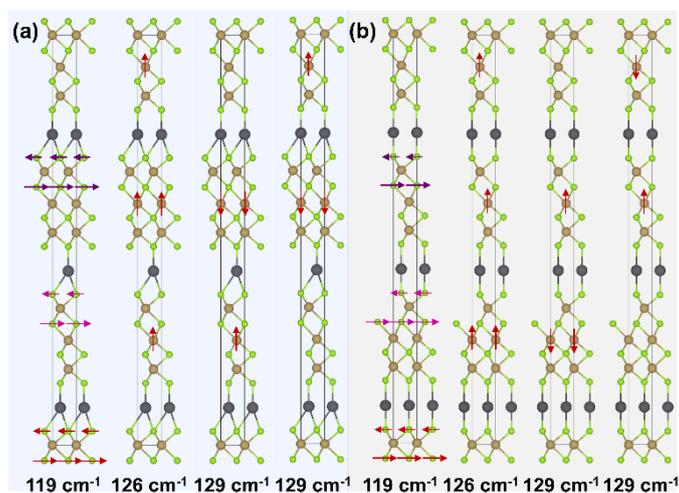

**Figure S3. The vibration patterns of Pb(Ta$_{1+x}$Se$_2$)$_2$ for phases I and III.** (a-b) The effects of the interstitial Ta atoms on the phonon frequencies for phase I and phase III.

It is found that there are two Raman peaks in the low frequency region caused by in-plane vibrations of Pb. We attribute the reason to the interference limitation of the laser peak, so it is not directly observed experimentally.

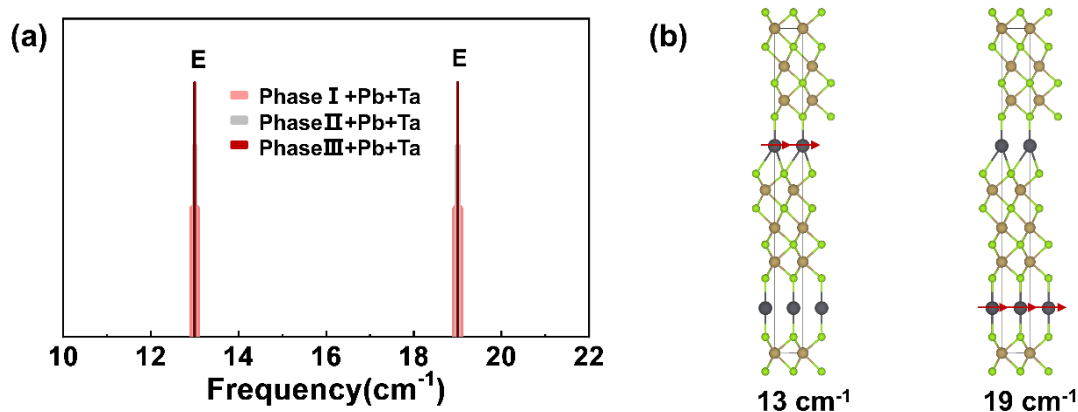

**Figure S4. DFT calculation of Pb(Ta$_{1+x}$Se$_2$)$_2$ Raman spectra for phase II.** (a) Different phonon modes within the 10-22 cm$^{-1}$ region. (b) The effects of the Pb atoms on the phonon frequencies of phase II in (a).

**References**

[1]  X. Yang, J.-K. Bao, Z. Lou, P. Li, C. Jiang, J. Wang, T. Sun, Y. Liu, W. Guo, S. Ramakrishnan, S. R. Kotla, M. Tolkiehn, C. Paulmann, G.-H. Cao, Y. Nie, W. Li, Y. Liu, S. van Smaalen, X. Lin, Z.-A. Xu. *Advanced Materials*. **2022**, 34, 2108550.
[2]  G. Kresse, J. Hafner. *Phys. Rev. B*. **1993**, 47, 558-561.
[3]  G. Kresse, J. Hafner. *Phys. Rev. B*. **1994**, 49, 14251-14269.
[4]  G. Kresse, D. Joubert. *Phys. Rev. B*. **1999**, 59, 1758-1775.
[5]  S. Grimme, S. Ehrlich, L. Goerigk. *J. Comput. Chem.* **2011**, 32, 1456-1465.
[6]  S. Baroni, S. de Gironcoli, A. Dal Corso, P. Giannozzi. *Reviews of Modern Physics*. **2001**, 73, 515-562.
[7]  M. I. Aroyo, A. Kirov, C. Capillas, J. M. Perez-Mato, H. Wondratschek. *Acta Crystallographica Section A*. **2006**, 62, 115-128.